\documentclass[aps,prb,twocolumn,superscriptaddress,10pt]{revtex4-2}
\usepackage{silence}

\usepackage{dcolumn}% Align table columns on decimal point
\usepackage{bm}% bold math
\usepackage{dsfont}%For integer symbol
\usepackage{epsfig}
\usepackage{amssymb} %additional symbol sets
\usepackage{amsfonts} %additional fonts, useful for mathematical symbols
\usepackage{mathtools} %improves equation appearance
\usepackage{dcolumn} %makes columns in tables align by decimal point
\usepackage{graphicx} %allows inclusion of graphics
\usepackage{color}  %introduces colors ("monochrome" feature useful for BW preview)
\usepackage{bm}  %allows bolding of mathematical symbols
\usepackage{comment} %allows large blocks of commenting
\usepackage{units}%allows \nicefrac to be used
\usepackage{textcomp} %gives a \textdegree command
\usepackage{units}%allows \nicefrac to be used
\usepackage{mathrsfs} 
\usepackage{array} %allows text in table to be vertically centered
\usepackage{natbib}
\usepackage{upgreek}
\usepackage{tabularx}
\usepackage{latexsym}
\usepackage{amsmath}
\usepackage{braket}
\usepackage{comment}
\usepackage{float}
\usepackage{bigstrut}
\usepackage{placeins}
\usepackage[dvipsnames]{xcolor}
\usepackage[breaklinks=true,pdfborder={0 0 0}]{hyperref}

\def \lsfo{La$_{1/3}$Sr$_{2/3}$FeO$_{3}$}
\def \sto{SrTiO$_3$}

\def \onesixth{$\left(\frac{1}{6} \textrm{ } \frac{1}{6} \textrm{ } \frac{1}{6}\right)$}
\def \onethird{$\left(\frac{1}{3} \textrm{ } \frac{1}{3} \textrm{ } \frac{1}{3}\right)$}

\def \onefourth{$\left(\frac{1}{4} \textrm{ } \frac{1}{4} \textrm{ } \frac{1}{4}\right)$}

\begin{document}
\title{Low-Temperature Suppression of Intertwined Orders in \lsfo\ Thin Films}
\author{Mayia A. Vranas}\email{These authors contributed equally}
\affiliation{Department of Physics, University of California San Diego, La Jolla, California 92093, USA}

\author{Robin Glefke}\email{These authors contributed equally}
\affiliation{Department of Physics, University of California San Diego, La Jolla, California 92093, USA}

\author{Katherine Matthews}\email{These authors contributed equally}
\affiliation{Department of Physics, University of California San Diego, La Jolla, California 92093, USA}

\author{Xiaoke Li}\email{These authors contributed equally}
\affiliation{Department of Physics, University of California San Diego, La Jolla, California 92093, USA}
\affiliation{Program in Materials Science and Engineering, University of California, San Diego, CA 92093, USA}

\author{Tianxing Wang}
\affiliation{Department of Physics, University of California San Diego, La Jolla, California 92093, USA}

\author{Henry Navarro}
\affiliation{Department of Physics, University of California San Diego, La Jolla, California 92093, USA}
\affiliation{Department of Physics, Andrews University, Berrien Springs, Michigan 491004, US}

\author{I-Ching Lin}
\affiliation{Condensed Matter Physics and Materials Science Division, Brookhaven National Laboratory, Upton, NY, USA}

\author{Sarmistha Das}
\affiliation{Department of Physics, University of California San Diego, La Jolla, California 92093, USA}

\author{Biswajit Sahoo}
\affiliation{Department of Physics, University of California San Diego, La Jolla, California 92093, USA}
\affiliation{Center for Memory and Recording Research, University of California San Diego, La Jolla, California 92093, USA}

\author{Rourav Basak}
\affiliation{Department of Physics, University of California San Diego, La Jolla, California 92093, USA}

\author{Wei He}
\affiliation{Department of Physics, University of California San Diego, La Jolla, California 92093, USA}
\affiliation{Advanced Light Source, Lawrence Berkeley National Laboratory, Berkeley, California 94720, USA}

\author{Holden Bauer}
\affiliation{Department of Physics, University of California San Diego, La Jolla, California 92093, USA}

\author{Ella Di Liberti}
\affiliation{Department of Physics, University of California San Diego, La Jolla, California 92093, USA}

\author{Jacob Butler}
\affiliation{Department of Physics, University of California San Diego, La Jolla, California 92093, USA}

\author{Elliot Kisiel}
\affiliation{Department of Physics, University of California San Diego, La Jolla, California 92093, USA}
\affiliation{X-ray Science Division, Argonne National Laboratory, Lemont, IL, 60439, USA}

\author{Erik Lamb}
\affiliation{Department of Physics, University of California San Diego, La Jolla, California 92093, USA}

\author{Jun-Sik Lee}
\affiliation{Stanford Synchrotron Radiation Lightsource, SLAC National Accelerator Laboratory, Menlo Park, California 94025, USA}

\author{Cheng-Tai Kuo}
\affiliation{Stanford Synchrotron Radiation Lightsource, SLAC National Accelerator Laboratory, Menlo Park, California 94025, USA}

\author{Heemin Lee}
\affiliation{Stanford Synchrotron Radiation Lightsource, SLAC National Accelerator Laboratory, Menlo Park, California 94025, USA}

\author{Christie Nelson}
\affiliation{National Synchrotron Light Source II, Brookhaven National Laboratory, Upton, New York 11973, USA}

\author{Sophie Morley}
\affiliation{Advanced Light Source, Lawrence Berkeley National Laboratory, Berkeley, California 94720, USA}

\author{Sujoy Roy}
\affiliation{Advanced Light Source, Lawrence Berkeley National Laboratory, Berkeley, California 94720, USA}

\author{Christoph Klewe}
\affiliation{Advanced Light Source, Lawrence Berkeley National Laboratory, Berkeley, California 94720, USA}

\author{Fanny Rodolakis}
\affiliation{X-ray Science Division, Argonne National Laboratory, Lemont, IL, 60439, USA}

\author{Jessica McChesney}
\affiliation{Argonne National Laboratory, Lemont, IL, 60439, USA}

\author{Oleg Shpyrko}
\affiliation{Department of Physics, University of California San Diego, La Jolla, California 92093, USA} 

\author{Yimei Zhu}
\affiliation{Condensed Matter Physics and Materials Science Division, Brookhaven National Laboratory, Upton, NY, USA}

\author{Eric Fullerton}
\affiliation{Center for Memory and Recording Research, University of California San Diego, La Jolla, California 92093, USA}

\author{Ivan Schuller}
\affiliation{Department of Physics, University of California San Diego, La Jolla, California 92093, USA}

\author{Alex Frano}\email{Corresponding author, email: afrano@ucsd.edu}
\affiliation{Department of Physics, University of California San Diego, La Jolla, California 92093, USA}
\affiliation{Program in Materials Science and Engineering, University of California, San Diego, CA 92093, USA}

\date{\today}

\begin{abstract}
The strong coupling between spin, charge, and lattice degrees of freedom in perovskite oxides leads to an array of exotic phenomena, giving these materials rich phase diagrams that can include coupled orders. This is exemplified by the A-site doped ferrite \lsfo\ (LSFO), which exhibits a coupled paramagnetic-antiferromagnetic and charge ordering phase transition at $\sim$190 K that has been well studied in thin films, bulk, and polycrystalline samples. However, the low temperature behavior of LSFO thin films below $\sim$100 K has not been thoroughly explored. This work uses several X-ray scattering and spectroscopy techniques to directly probe LSFO's magnetic and charge order down to low temperature. Using resonant X-ray scattering, we observe a complete suppression of LSFO's known antiferromagnetic and charge order below $\sim$25 K. Further spectroscopy and coherent scattering measurements provide insight into LSFO's electronic structure and domain dynamics in this new low temperature phase, and we propose possible explanations for the observed order suppression based on reduced dimensionality of domains in our thin films. Our findings provide insight into the effects of competing interactions in strongly correlated materials, particularly those with coupled orders.
\end{abstract}

\maketitle

\section{Introduction}
\par Transition metal perovskite oxides (ABO$_3$) are exciting systems of study due to their strong correlations and, as a result, their diverse material properties. Coupling between charge, spin, lattice, and orbital degrees of freedom can lead to high-temperature superconductivity \cite{bednorz1986possible}, colossal magnetoresistance \cite{ibarra1998colossal}, and metal-to-insulator transitions \cite{torrance1992systematic}. Alongside these exciting properties, these materials can also exhibit co-occurring or competing orders \cite{garcia1994neutron,taillefer2010scattering}, making them rich systems with multiple energy scales and interactions to consider.  

\par
\begin{figure*}[ht]
  \centering
  \includegraphics[width=\textwidth]{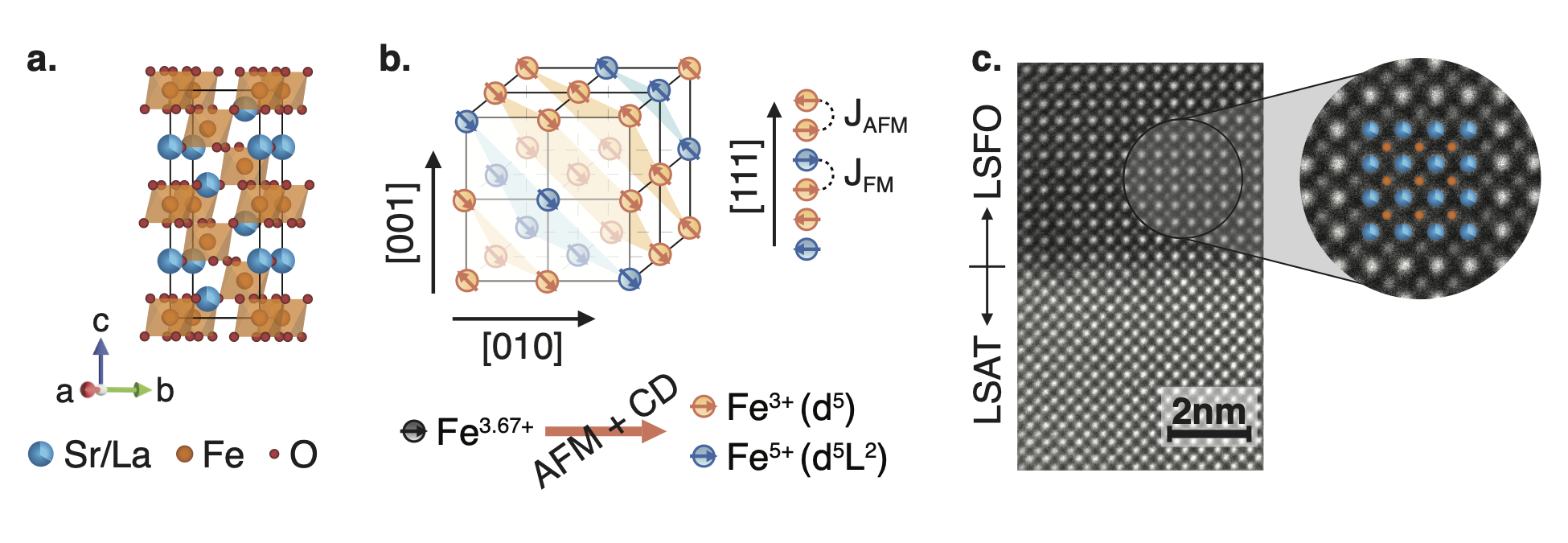}
  \caption{\textbf{LSFO Structure and Order.} a) Crystal structure of LSFO, modified from \cite{yangChargeDisproportionationOrdering2003}. b) Schematic of the CD (indicated by cation color) and AFM order (indicated by arrow direction) in LSFO. c) Transmission electron microscopy (TEM) image of a cross-section of LSFO grown on an LSAT(100) substrate. The Sr/La and Fe cation sites are labeled with blue and orange circles, respectively.}
  \label{fig:structure}
\end{figure*}

 Within the perovskite family, the mixed valence ferrite \lsfo\ (LSFO, Fig. \ref{fig:structure}) displays atypical ordering behavior. It exhibits antiferromagnetic (AFM) order ($T_N\sim190$K in thin films) with a remarkably long six-fold periodicity ($\uparrow\uparrow\uparrow\downarrow\downarrow\downarrow$) concomitant with charge disproportionation (CD) to an unusually high iron valence state (3Fe$^{3.67+}\rightarrow$ 2Fe$^{3+}$ $+$ Fe$^{5+}$)\cite{battleStudyChargeDisproportionation1988}. Unlike the many perovskites whose charge disproportionation arises from Jahn-Teller distortions \cite{kugel1982jahn}, the coupled orders in LSFO owe their stability to the relative strength of the ferromagnetic and antiferromagnetic interactions between Fe$^{3+}$-Fe$^{5+}$\  ($J_{FM}$) and Fe$^{3+}$-Fe$^{3+}$\  ($J_{AFM}$), respectively. The ferromagnetic coupling is a direct result of the large negative charge-transfer energy between the iron and oxygen, localizing doped holes in the oxygen $2p$ ligands \cite{abbateControlledvalenceProperties}, which hybridize with Fe $3d$ orbitals to create an effective Fe$^{5+}$\ valence state. The holes will prefer to sit in sites between Fe with the same spin direction, producing a ferromagnetic coupling between Fe$^{3+}$ and Fe$^{5+}$ sites. This hybridization allows for the formation of a high Fe$^{5+}$\ valence state without a Jahn-Teller distortion, as well as a stable charge-disproportionated phase driven primarily by magnetic exchange coupling (Fig. \ref{fig:structure}b)\cite{parkVariationChargeorderingTransitions1999,mcqueeneyStabilizationChargeOrdering2007}.

\par The vital role that LSFO's magnetic interactions play in stabilizing its CD emphasizes that these two orders are inherently and fundamentally coupled. These coupled orders have been studied in both bulk and thin film configurations. Both orders onset at the same temperature ($\sim$190 K) and form along the [111]$_{\textrm{pc}}$ (pc = pseudocubic) direction, with the $\uparrow\uparrow\uparrow\downarrow\downarrow\downarrow$ AFM order corresponding to two repetitions of the Fe$^{3+}$-Fe$^{5+}$-Fe$^{3+}$ charge order due to the Fe$^{3+}$-Fe$^{5+}$\ ferromagnetic and Fe$^{3+}$-Fe$^{3+}$\ antiferromagnetic exchange interactions described previously (Fig. \ref{fig:structure}b). LSFO's AFM and CD are robust in thin films with thicknesses down to 12nm \cite{minoharaThicknessdependentPhysicalProperties2016a,yamamotoThicknessDependenceDimensionality2018}. These orders occur without any significant long-range structural distortion \cite{battleStructuralConsequencesCharge1990}, though changes in the local bond environment have been observed \cite{herreromartinEvidenceChargedensitywaveNature2009, blascoChargeDisproportionation$mathrmLa_1ensuremathxmathrmSr_xmathrmFemathrmO_3$2008}. Accompanying the ordered state is a sharp upturn in resistivity (Fig. S1b in the Supplementary Information), though the system remains insulating on either side of $T_N$ despite a high density of charge carriers \cite{devlinElectronicTransportConduction2014}.

\par Despite extensive work in understanding the mechanisms surrounding the CD and AFM transitions, the low-temperature behavior of LSFO has not been thoroughly explored, especially in thin films. While initial evaluation would suggest that the ordered state should remain stable at low temperatures, LSFO's parent compound, SrFeO$_3$ (SFO), exhibits a wide array of competing magnetic structures that vary with temperature and applied field \cite{ishiwataVersatileHelimagneticPhases2011,reehuisNeutronDiffractionStudy2012,onoseCompletePhaseDiagram2020}. Likewise, low-temperature anomalies have been observed in LSFO; for example, an upturn is seen in the spontaneous magnetization at low temperatures in bulk, indicating that spins are aligning via canting to produce a non-zero magnetization \cite{parkVariationChargeorderingTransitions1999, treves1965studies}. The presence of spin canting indicates that there are additional coupling mechanisms at play, warranting a detailed study into the low-temperature behavior and the low-dimensional ordered ground state in LSFO thin films.

\par Here, we present a thorough investigation of the temperature-dependent behavior of the intertwined magnetic and charge orders in LSFO thin films. We report a novel phase transition discovered at low temperatures, suggesting an emerging competing phase or low-temperature disorder. Using resonant elastic X-ray scattering techniques in the soft and hard X-ray regimes (RSXS and REXS, respectively), we directly probe the behavior of the $\vec{q}=$\onesixth\ AFM ordering peak at Fe $L_3$ edge and the $\vec{q}=$\onethird\ CD ordering peak at Fe $K$ edge. Remarkably, we observe that both peaks diminish at $T^*\sim60$ K and ultimately disappear at $T_{supp}=20$ K in films of varying thickness, orientation, and substrate, indicating a collapse of the long-range order and suggesting that the ground state of LSFO is more complicated than the canonical high-temperature model. Moreover, our temporal-scale dynamical measurements using X-ray Photon Correlation Spectroscopy (XPCS) reveal that the dynamics associated with the AFM order in LSFO accelerate when moving through the low-temperature melting transition. These results establish LSFO as a model system for studying phase competition in transition-metal oxides while highlighting the strength of coherent scattering techniques to disentangle static and dynamic components of ordering phenomena in strongly correlated systems.

\begin{figure*}[ht]
  \centering
  \includegraphics[width=\textwidth]{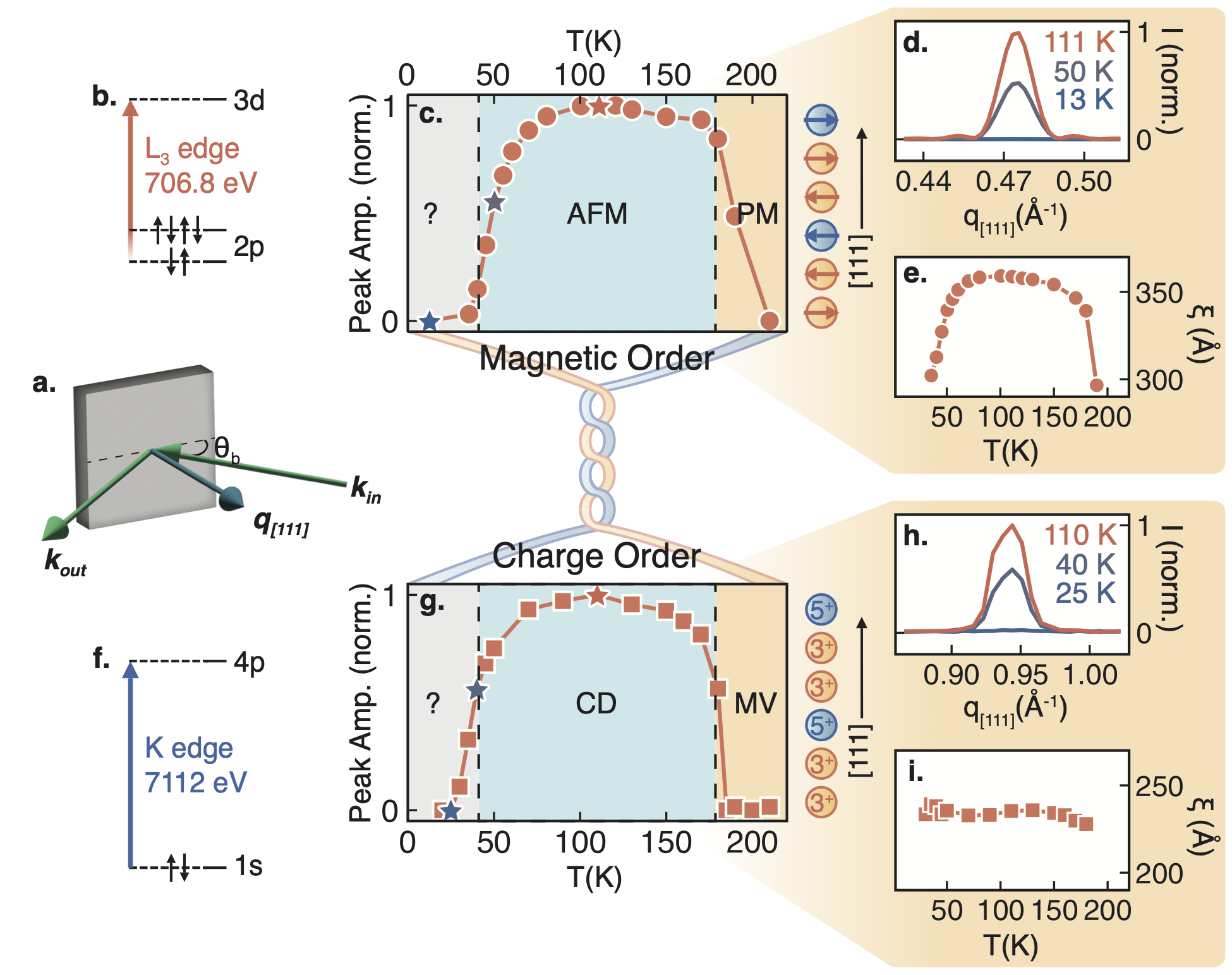}
    % \caption*{HKL Scan}
  \caption{\textbf{Resonant X-Ray Scattering Results.} a) Horizontal scattering geometry of our soft resonant X-ray scattering (RSXS) experiments. Hard resonant X-ray scattering (REXS) was done in a vertical scattering geometry instead. b,f) Energy level diagrams of the electronic excitations for our Fe $L_3$ edge RSXS and Fe $K$ edge REXS measurements, respectively. c,g) Temperature dependence of the amplitudes of LSFO's AFM and CD Bragg peaks, $\vec{q}=$\onesixth\ and $\vec{q}=$\onethird\ , respectively. Regions dominated by paramagnetic (PM) mixed valence (MV), antiferromagnetic (AFM) charge disproportionated (CD), and unknown low-temperature phases (?) are delineated. Colored stars correspond to curves in panels d,h), which show example AFM and CD peak profiles. e,i) Temperature dependence of the out-of-plane correlation lengths of the AFM and CD peaks, respectively.}
  \label{fig:result}
\end{figure*}

\begin{figure*}[ht]
  \centering
  \includegraphics[width=\textwidth]{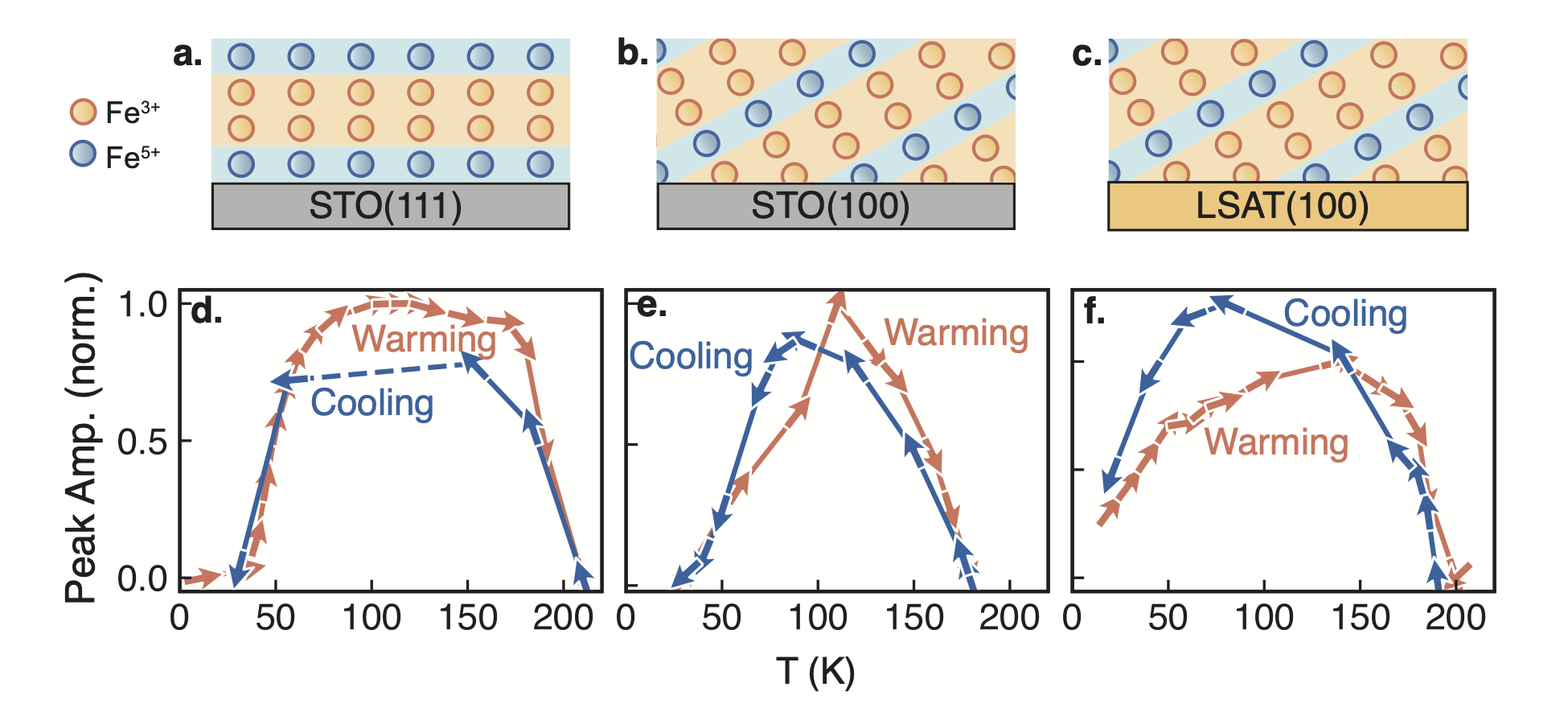}
  \caption{\textbf{Substrate Comparison.} Temperature dependence of the $\vec{q}=$\onesixth\ magnetic peak for LSFO films grown on a,d) (111)-oriented SrTiO$_3$, b,e) (100)-oriented SrTiO$_3$, and c,f) (100)-oriented LSAT substrates.}
  \label{fig:hysteresis}
\end{figure*}

\section{Results}
\par The low-temperature behaviors of the $\vec{q}=$\onesixth\ AFM peak and the $\vec{q}=$\onethird\ CD peak are characterized for 46 nm and 29 nm films, respectively, grown on (111)-oriented \sto (STO) substrates (Fig. \ref{fig:result}). Both peaks display the expected upturn in intensity through $T_N$, which stabilizes by $T=100$ K. However, both peaks later experience a steep decrease at $T^*\sim 60$ K, with complete suppression seen at $T_{supp}\sim20$ K (Fig. \ref{fig:result}c,g). Small variation is seen in the values of $T_N$, $T^*$, and $T_{supp}$ due to sample thickness variation. The energy dependence of the scattering in the AFM and the CD peaks shows a resonant enhancement around the Fe $L_3$ and $K$ edges respectively, confirming that these reflections are contributed to by the Fe sites (Fig. S2). From fits of the peak profiles along the [111]$_{\textrm{pc}}$ direction (Fig. \ref{fig:result}d,h), correlation lengths of the magnetic and charge orders were extracted, and found to be 357 Å and 234 Å, respectively (Fig. \ref{fig:result}e,i). These correlation lengths are similar to the thicknesses of the films ($\sim$80\% of the film thickness in both cases), indicating that magnetic and CD correlations are robust throughout the film and confirming that the orders are long-range rather than local.

\par The observation of this low-temperature suppression of long-range order in thin films was not seen previously in single-crystalline LSFO, raising the question of whether it originates from substrate effects, especially given that STO experiences a structural transition at $T_S=100$ K \cite{okazaki1983time, shirane1969lattice}. To determine whether this suppression was growth orientation- and/or substrate-specific, further RSXS experiments were conducted on the $\vec{q}=$\onesixth\ magnetic peak for films grown on STO(100) and LSAT(100) in the same growth conditions. The same suppression of order was seen for all films, with some variation in $T_N$ and $T^*$ (Fig. \ref{fig:hysteresis}). LSFO films also appear to be well-ordered in chemical structure without visible displacement defects in TEM (Fig. \ref{fig:structure}c). Therefore, the observed order suppression is intrinsic to the LSFO film and is not due to substrate influences or structural distortion. 

With the AFM and CD suppression in LSFO consistently observed across multiple samples, substrates, orientations, and thicknesses, this unexpected but highly reproducible result raises the question of what could be driving the disappearance of the known orders and what novel phase(s) take their place. Our resonant scattering measurements already provide a few hints; hysteretic behavior was observed in the magnetic peak's temperature dependence for films on all substrates. This may indicate first-order nature in the low-temperature transition and is in agreement with previous results indicating that the high-temperature transition at $T_N$ is first order \cite{liChargeOrderedStates1997,mcqueeneyStabilizationChargeOrdering2007,herreromartinEvidenceChargedensitywaveNature2009,zhu2018unconventional}. If the low temperature transition is indeed first order, it will involve domain nucleation, interactions, and phase coexistence within a particular temperature regime. Further, the correlation length analysis of the two orders reveals that the magnetic order correlation length changes more significantly in a wide temperature range around each transition temperature. Similar to the magnetic peak amplitude, the correlation length of the AFM order changes as a function of temperature, increasing below $T_N$ and decreasing through the melting transition (Fig. \ref{fig:result}e). In contrast, the correlation length of the CD order remains relatively constant between the two phase transitions (Fig. \ref{fig:result}i). Together, the first order nature of the transition and the temperature sensitivity of the magnetic correlation length establish significant changes in LSFO's magnetic domains leading up to the low temperature suppression.

\begin{figure*}[ht]
  \centering
  \includegraphics[width=\textwidth]{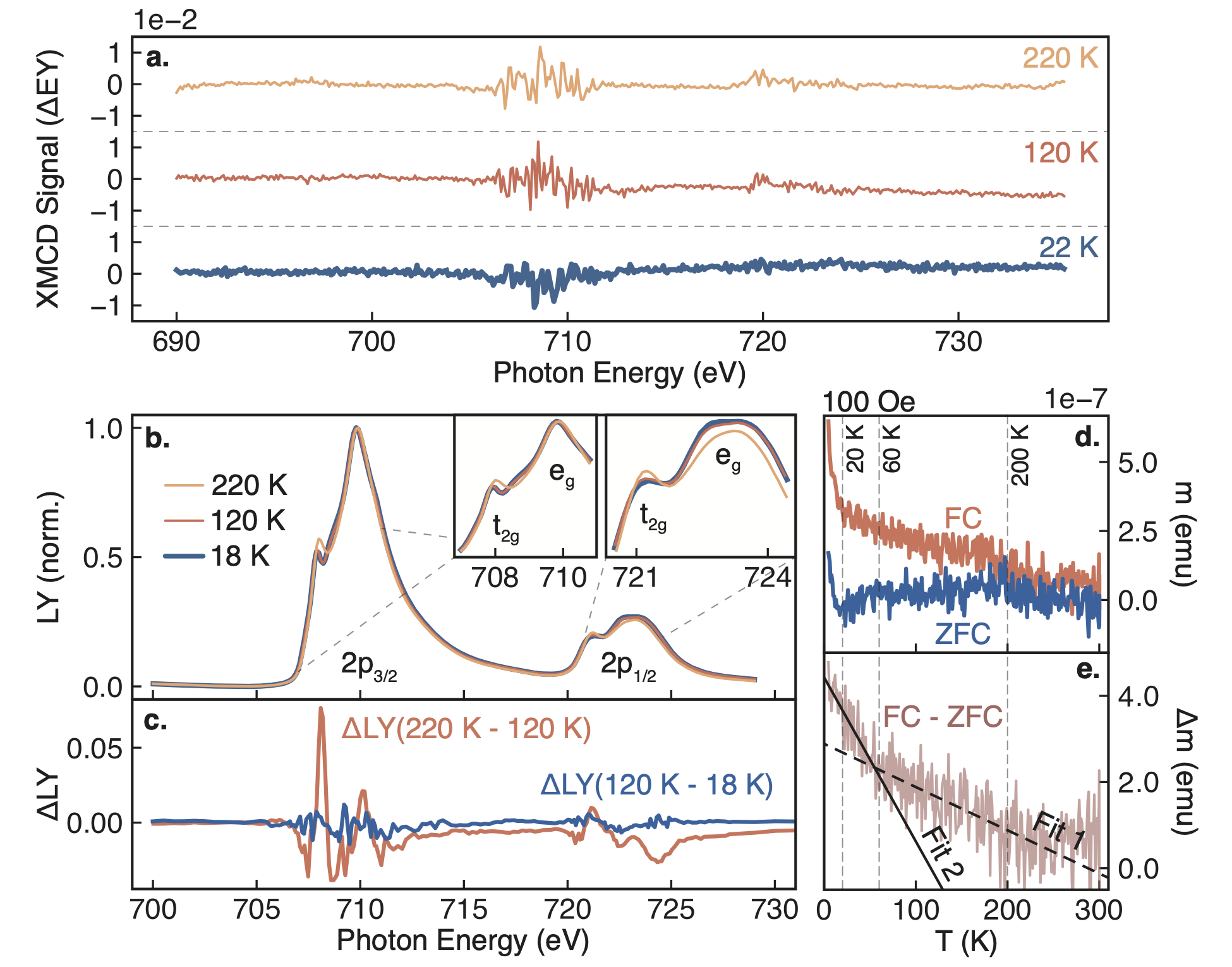}
  \caption{\textbf{Low Temperature Magnetic and Electronic Probes.} a) Fe $L_{2,3}$ edge XMCD measurements at three temperatures between 22 K and 220 K. b) Fe $L_{2,3}$ edge X-ray absorption spectroscopy (XAS) of LSFO at three temperatures taken in luminescence yield (LY). The $L$-edge peaks are labeled according to excited core state ($L_3$ - $2p_{3/2}$; $L_2$ - $2p_{1/2}$) and final crystal field split valence states (3d $e_g$ and $t_{2g}$)\cite{abbateControlledvalenceProperties,braunSpectralIndicatorHoleDepletion2009}. c) The difference in absorption spectra from 220 K to 120 K (orange) and from 120 K to 18 K (blue), highlighting the changes in spectral weight across both transition temperatures. d) Field-cooled (FC) and zero field-cooled (ZFC) moment vs temperature measurements of LSFO. e) Difference between FC and ZFC measurements with 2 linear fits, showing a change in slope at low temperatures.}
  \label{fig:FIG4}
\end{figure*}

\par The next natural hypothesis for the origin of this low-temperature transition is the formation of a new ordered state. Previous studies have suggested that multiple magnetic exchange interactions are possible in LSFO, such that a 1/4-period magnetic order could emerge \cite{parkSuperlatticeinducedFerroelectricityChargeordered2019}. A search for magnetic peaks at the Fe $L_3$ and $L_2$ edges along the  [111]$_{\textrm{pc}}$ direction of an LSFO sample grown on STO(111) did not reveal any new magnetic propagation vectors, including $\vec{q}=$\onefourth\ (Fig. S3a-b). However, access to reciprocal space is limited in the soft X-ray regime, severely limiting the breadth of our search. It is also possible that a new magnetic order forms along a different direction in reciprocal space that was not probed in our search. 

\par Indications of a new ordered state may also appear in the material's X-ray absorption spectra (XAS). The 1/6-period magnetic order seen at intermediate temperatures requires the charge ordering of Fe$^{3+}$\ and Fe$^{5+}$\ ions, so that the appropriate exchange couplings $J_{FM,AFM}$ can mediate the long-range order (Fig. \ref{fig:structure}b). A new ordered state may require a different change in the Fe valence, which would appear as a shift in spectral weight in the Fe edge XAS. Indeed, through $T_N$, a small shift in spectral weight is seen in the Fe $L_{2,3}$ edges (Fig. \ref{fig:FIG4}b, c), corresponding to electrons moving from the 3d $t_{2g}$ to $e_{g}$ orbitals with the onset of the known charge disproportionation \cite{abbateControlledvalenceProperties,higuchi2011electronic}. However, no significant shift in spectral weight is seen between 120K and 18K (Fig. \ref{fig:FIG4}c), indicating that there is no change in Fe valence through $T^*$. Additionally, electrical resistance measurements (Fig. S1b) do not show a clear indication of this low temperature transition like the upturn at T$_N$, which could reflect the lack of change in electronic structure at low temperature, in agreement with the absorption spectra.
\par Though no additional AFM ordering peak is observed, the possibility of a new ferromagnetic (FM) order cannot be ruled out using RSXS experiments. This is because the corresponding FM Bragg peak would overlap with, and therefore be indistinguishable from, the structural Bragg peaks. Furthermore, in RSXS, limited reciprocal space access at the Fe $L_{2,3}$ edges precludes observation of structural peaks, which lay outside the Ewald sphere at these energies. Previous results have suggested the presence of spin canting in LSFO, indicating that FM coupling may be significant \cite{battleStructuralConsequencesCharge1990, parkVariationChargeorderingTransitions1999, li2002microstructure,dasilvaOriginSpinglassExchange2014, onoseCompletePhaseDiagram2020}. To explore the possibility that ferromagnetic coupling begins to dominate at low temperatures, X-ray magnetic circular dichroism (XMCD) measurements were performed. This technique takes the difference in absorption signal of left- and right-circularly polarized X-rays, with large XMCD signals indicating ferromagnetism in a material. As shown in Fig. \ref{fig:FIG4}a, the measured Fe L$_{2,3}$ edge XMCD in LSFO is negligible at all temperatures, indicating that a ferromagnetic phase is not present at low temperatures.
\par To further probe LSFO's magnetic behavior, vibrating sample magnetometry (VSM) measurements were conducted on LSFO thin films in both field-cooled (FC) and zero-field cooled (ZFC) conditions (Fig. \ref{fig:FIG4}d). The sample was cooled to base temperature (5K) with (FC) or without (ZFC) a 100 Oe magnetic field applied and then was measured with a 100 Oe field while warming to room temperature. A sharp splitting of the FC-ZFC magnetization is observed at $T_N$, which has been observed previously in this material and is attributed to the onset of spin canting concomitant with the AFM order \cite{battleStructuralConsequencesCharge1990, parkVariationChargeorderingTransitions1999, li2002microstructure,dasilvaOriginSpinglassExchange2014, onoseCompletePhaseDiagram2020}. At $T^*\sim60$ K, the magnitude of the FC-ZFC splitting increases, as reflected by an increase in the slope of $\Delta$FC-ZFC by a factor of 3.8 (Fig. \ref{fig:FIG4}e). At $T_{supp}\sim20$ K, an upturn in both the FC and ZFC total moments is seen. These results align well with the relevant temperatures in scattering, where at $T\sim60$ K the magnetic order suppression begins, and at $T_{supp}\sim20$ K, the magnetic order is completely suppressed. FC-ZFC splitting can be indicative of a canted AFM order or the onset of a spin glass \cite{kumarCantingZFCIrreversability2013,chunSpinGlassCantedAFM2001}. Spin glasses and related phases tend to emerge when there is competition between different interactions combined with disorder in the system and are typically associated with a slowdown of magnetic dynamics in the system. 

\begin{figure*}[ht]
  \centering
  \includegraphics[width=\textwidth]{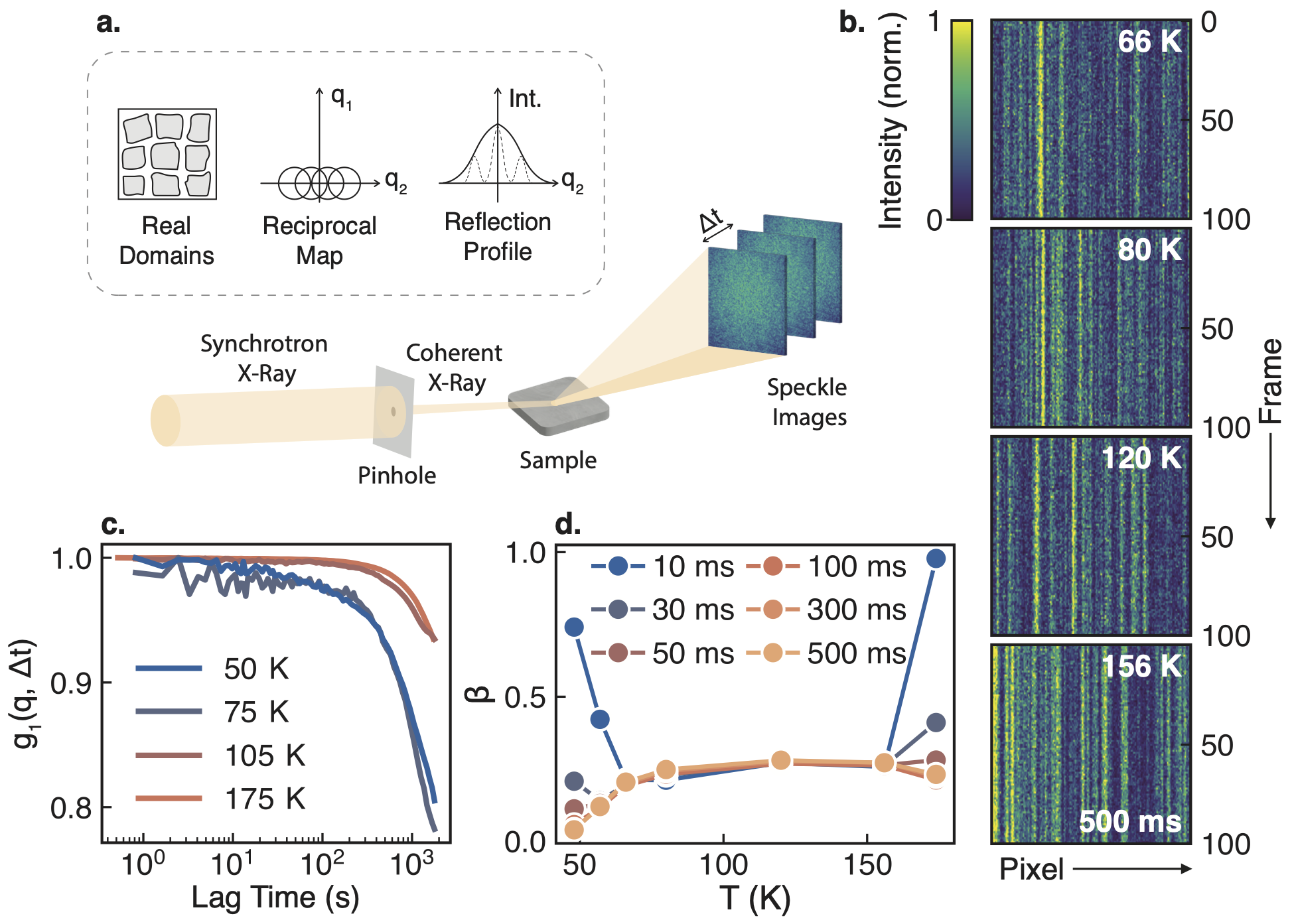}
  \caption{\textbf{X-ray Photon Correlation Spectroscopy (XPCS) of the AFM Peak.} a) A schematic of the coherent scattering process, in which a coherent beam scattering off of magnetic domains in the sample creates a speckle pattern that varies over time. b) Waterfall plots for four temperatures, where a line cut across the Bragg peak is plotted for every frame recorded during measurement, showing the evolution of speckled in time. c) The intermediate scattering function, $g_1$, plotted for four temperatures. The low temperature and high temperature dynamics show different timescales. d) Temperature-dependent contrast of speckles for six exposure times, showing an increase in contrast at low temperatures. }
  \label{fig:FIG5}
\end{figure*}

\par The possibility of a spin glass or a frustrated state warrants investigation of low-temperature dynamics through $T^*$ in LSFO thin films; however, typical markers of a spin glass, such as a shift in $T_g$ at different frequencies in AC susceptibility, are difficult to obtain with small sample volumes. We instead investigated the dynamical behavior of LSFO using X-ray Photon Correlation Spectroscopy (XPCS) on the $\vec{q}=$\onesixth\ AFM ordering peak. This technique utilizes a coherent incident X-ray beam to resolve small variations in the scattering of different magnetic domains, resulting in a speckle pattern that reflects the domain size and dynamics (Fig. \ref{fig:FIG5}a). Fluctuations over time in these speckles directly correspond to domain fluctuations, and the timescale of domain dynamics, $\tau$, can be extrapolated from the intermediate scattering function: 
\begin{equation}
    g_1(q,\Delta t)=e^{-(\Delta t/\tau)^\gamma}.
\end{equation}
The $g_1$ quantifies fluctuations to the electric field over a time $\Delta t$ and is therefore not directly measurable with this technique. Instead, the intensity autocorrelation function, $g_2$, is measured:
\begin{equation}
    g_2(q,\Delta t) = \frac{\braket{I(q,t)I(q,t+\Delta t)}}{\braket{I(q,t)}^2},
\end{equation}
where $I(q,t)$ and $I(q,t+\Delta t)$ are the speckle intensities at times $t$ and $t+\Delta t$, respectively. The $g_1$ is related to the $g_2$ by the  Siegert relation:
\begin{equation}
    g_2(q,\Delta t) = 1+\beta\,\left|g_1(q,\Delta t) \right|^2,
  \label{equation:siegert}
\end{equation}
where $\beta$ is the speckle visibility, ranging from 0 to 1 \cite{lehmkuhlerDynamicCoherentXrayMeasurements2021}. 

\par The $g_1$ extracted from the AFM ordering peak using the above equations exhibits different behavior in the high- and low-temperature regimes (Fig. \ref{fig:FIG5}c), with the faster-decaying exponential indicating faster dynamics at $T=50$ K than at a point through the AFM transition at $T=175$ K. Additionally, for $T=50$ K, the $g_1$ curve deviates from the horizontal at $t\sim10^1$ s before the sharp dropoff near $t\sim10^3$ s. This may suggest the existence of multiple timescales at $T=50$ K, which would result in a $g_1$ that is the sum of multiple exponential functions; it may also indicate dynamical timescales faster than the resolution of the detector. 

\par Waterfall plots (Fig. \ref{fig:FIG5}b) provide a visualization of the speckle pattern's fluctuations by plotting a slice of the measured Bragg peak over time, tracking the dynamics of individual speckles. Each column of pixels in the waterfall plots reflects the intensity variation of a single pixel through the plotted time duration, with each row shown corresponding to a frame with a 500 ms exposure time. The waterfall plots show a visible difference in contrast as a function of temperature, even when normalized to overall intensity; the columns in the 66 K waterfall plot look less sharp or distinct than those in the 80 K plot. This indicates that the motion of the speckles is faster than the 500 ms exposure time at 66 K, leading to ``blurred'' features.

\par To estimate possible dynamics faster than the detector resolution, we look to the speckle visibility $\beta$ as a function of exposure time and temperature, which quantifies speckle blurring (Fig. \ref{fig:FIG5}d). For static behavior, $\beta$ will not change with varying exposure times, as seen in the AFM region $80\mathrm{K}<T<156\mathrm{K}$; speckles do not fluctuate in intensity in a static regime and therefore are not impacted by the exposure time in this regime. However, through $T_N$ and $T^*$, a change in speckle visibility $\beta$ is seen with exposure time, supporting the statement that these regimes are dynamical. 

\section{Discussion}

\par In summary, our measurements reveal a low-temperature melting of CD and AFM order in epitaxial LSFO thin films. This melting transition is observed for films grown on STO and LSAT substrates and for different growth orientations, indicating that this behavior is intrinsic to LSFO. This novel disappearance of LSFO's known orders at low temperature motivates deeper considerations of the system's intrinsic coupling and competing energy scales, particularly in LSFO thin films. 

\par The coupling between the magnetic and charge orders in LSFO is central to the material's fundamental properties and ordering. LSFO has doped holes that prefer to localize on the oxygen ligands, leading to hybridization of oxygen $2p$ orbitals with iron $3d$ orbitals and creating an effective Fe$^{5+}$\ valence. These sites exhibit a ferromagnetic exchange with the neighboring Fe$^{3+}$\ ions mediated by the oxygen ligands, in contrast with the antiferromagnetic exchange between the Fe$^{3+}$\ sites. The strength of the Fe$^{3+}$-Fe$^{5+}$\ ferromagnetic coupling, $J_{FM}$, has been confirmed to be greater than the strength of the Fe$^{3+}$-Fe$^{3+}$\ antiferromagnetic coupling $J_{AFM}$, such that $|J_{FM}/J_{AFM}|=1.5$ \cite{mcqueeneyStabilizationChargeOrdering2007}. This experimentally derived relative coupling strength is in agreement with a charge ordered state driven primarily by magnetic interactions \cite{mizokawaDescriptionSpinCharge1998}. Therefore, if the magnetic order changes or becomes destabilized, the charge order likely will too, as we observed.  

\par While the magnetic and charge orders are inherently related and are coupled in both their onset at $T_N$ and their suppression below $T_{supp}$, their behaviors in proximity to both phase transitions differ significantly. Correlation length analysis of our magnetic and charge scattering experiments, shown in Fig. \ref{fig:result}e,i, reveals that the out-of-plane CD correlation length remains stable for the entire temperature range where there is a measurable peak. The AFM correlation length, however, decreases near the two transition temperatures $T_N$ and $T_{supp}$, showing a temperature profile similar to that of the magnetic peak amplitude shown in Fig. \ref{fig:result}c. Therefore, both the number (peak amplitude) and the size (correlation length) of the AFM ordered domains decrease near the transition temperatures, whereas the CD domains retain an approximately constant size while decreasing in number. This indicates that the thermal evolution of the two orders' domains are distinct, possibly due to their differing periods, and suggests that the magnetic order is more susceptible to thermal destabilization.

\par Distinct behavior of LSFO's magnetic and charge correlation lengths has been reported previously in a study of the thickness dependence of the AFM and CD behavior in LSFO thin films \cite{yamamotoThicknessDependenceDimensionality2018}. Using resonant X-ray scattering, Yamamoto et al. found that LSFO's CD correlation length remains relatively constant regardless of film thickness while its AFM correlation length scales with the film thickness \cite{yamamotoThicknessDependenceDimensionality2018}. At $\sim$ 15 nm, the magnetic correlation length is reduced to the scale of the charge order correlation length, and below this threshold both orders destabilize due to their strong interdependence. By directly probing LSFO's AFM and CD domains with X-rays, this study revealed that LSFO's AFM order is more sensitive to reduced dimensionality than its charge disproportionation and that there is a minimum thickness required for long-range order. 

\par Another study investigated dimensionality effects in nanocrystalline LSFO \cite{sabyasachiGlassyMagneticPhase2012}. Sabyasachi et al. found that smaller particle size ($\sim$70 nm) produces short-range CD and AFM order, while bulk material ($\sim$200 nm) retains the conventional long-range ordered state. Their Mössbauer analysis showed that the nanocrystalline sample does not undergo a change in spin state or a substantial change in the Fe$^{3+}$/Fe$^{5+}$ disproportionation relative to bulk; instead, the main effect of size reduction is to destabilize long-range order. Supported by AC susceptibility and exchange bias measurements, Sabyasachi et al. propose that their nanocrystalline LSFO exhibits a cluster glass-like phase resulting from the short-range order and multiple competing magnetic interactions at grain boundaries, leading to spin frustration. While this work differs from ours in the type of LSFO sample and the lack of any long-range order in their smallest nanocrystals, the vastly different behavior they observe in their 70 nm samples compared to their 200 nm bulk samples clearly demonstrates that finite size effects can strongly affect order formation and stability in LSFO. 

\par Placing our work in the context of these prior studies helps elucidate the mechanisms behind this low temperature suppression of LSFO's known order. The paper by Sabyasachi et al. established that dimensionality is a crucial factor in the formation and stability of charge and magnetic order in LSFO \cite{sabyasachiGlassyMagneticPhase2012}. However, this study focused on nanocrystalline LSFO with particle sizes between 70 nm and 200 nm, so it is not directly comparable to our $\sim$ 15-45 nm-thick LSFO thin films. Nevertheless, as discussed earlier, other studies have tested the thickness dependence of magnetic and charge order in LSFO thin films like ours. Using resonant X-ray scattering \cite{yamamotoThicknessDependenceDimensionality2018} and resistivity measurements \cite{minoharaThicknessdependentPhysicalProperties2016a}, they established that a minimum film thickness of 5-15 nm is required to support long-range order in LSFO. These studies confirmed that similar dimensionality effects are significant in thin films as well, but they only included temperature dependencies down to 120 K for bulk-sensitive wide angle scattering measurements \cite{yamamotoThicknessDependenceDimensionality2018} or 100 K for resistivity measurements \cite{minoharaThicknessdependentPhysicalProperties2016a}. These limited temperature ranges for the thin film studies exclude the low temperature regime where we observe the order suppression. Therefore, based on the results and gaps in the prior literature, we believe that the combination of the relatively small thickness of our films and the low temperature regime is key to this observed order suppression in LSFO.

\par We propose that in reduced dimensions, the long-range coupled CD/AFM order is susceptible to destabilization below $\sim 100$ K. LSFO's AFM order has a particularly long period\,---\,with one magnetic unit cell consisting of six Fe sites\,---\,and its correlation length has been shown to be particularly sensitive to thickness \cite{yamamotoThicknessDependenceDimensionality2018} and temperature (Fig. \ref{fig:result}e). While the charge disproportionation has a shorter period and is far less susceptible to thickness \cite{yamamotoThicknessDependenceDimensionality2018} and temperature (Fig. \ref{fig:result}i) changes, it is driven and stabilized by the more fragile magnetic order \cite{mcqueeneyStabilizationChargeOrdering2007, mizokawaDescriptionSpinCharge1998}. We suspect that our films are thick enough to host long-range order at intermediate temperatures but are close enough to the critical thickness to be susceptible to destabilization when thermal energy is reduced. Therefore, we propose that the low dimensionality and low temperature regime destabilizes the long period AFM order, which in turn destabilizes the coupled CD and results in the disappearance of the $\vec{q}=$\onesixth\ and $\vec{q}=$\onethird\ peaks below $T_{supp}$. 

\par Even with our postulated mechanism of the AFM and CD suppression, the question remains as to what low temperature phase is present in our LSFO thin films. Our measurements provide some insight: X-ray absorption spectroscopy reflects a change in spectral weight across $T_N$ with the onset of CD but no change in spectral weight across $T_{supp}$. This suggests that the 2:1 ratio of Fe$^{3+}$ to Fe$^{5+}$ is maintained despite the CD peak disappearing. This maintained valence ratio at low temperature is further reflected in our transport measurements (Fig. S1b), which do not show a low temperature kink like the one at $T_N$ corresponding to the onset of CD. No new magnetic ordering peak emerges when probed along the [111]$_{\textrm{pc}}$ direction (Fig. S3a-b), though this search is strongly limited by the small amount of reciprocal space accessible at the Fe $L_{2,3}$ edges. XMCD remains minimal in the low temperature regime, and from magnetometry, we see that below $T^*$ LSFO has an increased FC-ZFC split, possibly suggestive of spin canting or frustration. Lastly, from XPCS, we see that LSFO exhibits faster magnetic dynamics leading up to $T_{supp}$ compared to near $T_N$. In summary, we know the low temperature phase is not ferromagnetic, maintains the 2:1 ratio of Fe$^{3+}$ to Fe$^{5+}$ valence states, and exhibits FC-ZFC splitting with fast dynamics near the transition temperature. Two key candidates remain as the possible low temperature phase: 1) a lower energy AFM and/or CD ground state, or 2) a glassy state resulting from the destabilization of long-range order.

\subsection{A New AFM/CD Phase}
\par While our search along the [111]$_{\textrm{pc}}$ direction in RSXS measurements did not reveal any new AFM ordering peaks (Fig. S3a-b), the limited reciprocal space access available to us at the Fe $L_{2,3}$ edges means that we cannot eliminate the possibility of another AFM phase at low temperatures. An AFM phase with a periodicity inaccessible to us or along a direction other than [111]$_{\textrm{pc}}$ would have evaded our search. Spin canting within a new AFM phase could explain the enhanced FC-ZFC splitting seen in our magnetometry measurements. Additionally, given the coupling between the magnetic and charge orders at intermediate temperatures, a low temperature AFM phase could reasonably be accompanied by another CD phase. This continuity of spin order and charge disproportionation across $T_{supp}$ would be consistent with the lack of change in spectral weight observed in XAS measurements, which indicates that the system maintains its 2:1 Fe$^{3+}$:Fe$^{5+}$ ratio instead of reverting back into the averaged Fe$^{3.67+}$ state seen in its high temperature paramagnetic, mixed valence phase. Finally, the transition from the $\vec{q}=$\onesixth\ AFM and $\vec{q}=$\onethird\ CD phase into a different but still ordered AFM and CD state could be consistent with our proposed suppression mechanism described above; with the long period $\vec{q}=$\onesixth AFM order destabilized by the film's low dimensionality at low temperatures but the CD relatively unperturbed on its own, the system transitions into a lower energy (shorter period, different ordering direction or type, etc) AFM state, and the CD accompanies it due to its reliance on magnetic interactions for stability. 
\par A transition to a different AFM ordered state in LSFO would require an additional magnetic exchange interaction that pulls the spins into a different magnetic ground state. The presence of several different magnetic exchanges is indeed a hallmark of many iron perovskite materials, which can provide additional insight into the exchange interactions that may be present in LSFO. In the alkaline earth metal ferrites AFeO$_3$ (A=Ca, Sr, Ba), for example, A-site cations take on a $2+$ oxidation state, resulting in an unstable Fe$^{4+}$ valence \cite{clevenger1963}. A large, negative charge transfer in these materials results in an effective $d^5\underline{L}$ valence \cite{bocquet1992, hayashi2011,rogge2018}. In CaFeO$_3$, the iron sites undergo a further charge disproportionation of $2d^5\underline{L}\rightarrow d^5 + d^5\underline{L}^2$ (i.e. Fe$^{4+}\rightarrow\ $Fe$^{3+}+$Fe$^{5+}$) with a lattice distortion. As a result, electrons are localized; despite the FM exchange between these sites, it is theorized that superexchange between next-nearest neighbors via the oxygen ligands instead produces a helical or sinusoidal antiferromagnet \cite{takano1983,woodward2000,yang2005,mostovoy2005}. Conversely, in SrFeO$_3$ and BaFeO$_3$, electrons become delocalized, leading to a metallic conductivity. Differences in the strength of the nearest-neighbor ferromagnetic coupling and antiferromagnetic superexchange between these two materials and their cubic crystalline symmetry results in different helical antiferromagnetic orders \cite{hayashi2011,takegami2024,takano1983,andriushin2025}. In SrFeO$_3$, an even greater array of helical orders is present upon the application of magnetic field, while in BaFeO$_3$, which has a larger lattice parameter, magnetic field creates a ferromagnetic state. These cases highlight how nearest- and next-nearest neighbor interactions, lattice size, and charge localization can allow iron perovskites to take on multitudinous ferromagnetic and antiferromagnetic configurations.
\par Spin reorientation, meanwhile, is quite common for the rare-earth (RE) iron perovskites RFeO$_3$ (R=rare earth element) when the R-site cation is magnetic. In these materials, R takes on a $3+$ oxidation state, so that the iron sites are Fe$^{3+}$. As a result, at a $T_N$ that varies from $T_N=740$ K (R=La) to $T_N=623$ K (R=Lu) \cite{eibschutz1967} due to changes in the Fe-O-Fe bond angle \cite{lyubutin1999}, a G-type antiferromagnet will emerge \cite{koehler1957,seo2008,ritter2021,yuan2011,kuo2014,ritter2022}, which is sometimes canted \cite{tsymbal2007}. If R$^{3+}$ is magnetic, the interaction of the magnetic RE sublattice and the magnetic Fe sublattice competes with the magnetic exchange within the Fe sublattice, giving rise to a spin reorientation \cite{ritter2021,yuan2011,cao2014,tsymbal2007}. These reorientations do not change the position of the magnetic reflections, but would change their intensities as the spins rotate. Notably, in LaFeO$_3$, La$^{3+}$ is not magnetic, so this spin reorientation does not occur.
\par This survey of iron perovskite materials allows us to put the novel low-temperature phase of LSFO in better context. In particular, it is seen that a variety of magnetic structures can emerge in these materials, which are highly sensitive to the iron valence, charge-transfer energy of electrons from the iron sites to the oxygen ligands, structural distortion, and magnetism in the rare-earth or alkaline sublattice. LSFO possesses neither the magnetic rare-earth sublattice that produces spin reorientation in RFeO$_3$ materials, nor structural distortion as a driving force in its charge disproportionation as in CaFeO$_3$. Like in the alkaline iron perovskites, a large negative-charge transfer allows a high valence state to be stabilized; however, it does not possess the cubic symmetry responsible for the multi-q phases in SrFeO$_3$ and BaFeO$_3$, and instead finds a stable long-period magnetic configuration. In light of these differences, LSFO is not directly comparable to its perovskite cousins, implying a different mechanism for the low-temperature order suppression.
\par Currently, our studies do not rule out the possibility of a spin reorientation or reordering in LSFO. In the absence of a rare-earth magnetic sublattice, such a reorientation may require a structural distortion. Such a distortion could alter the ratio of the relevant energy scales, namely the charge-transfer energy $\Delta_{CT}$, the crystal field splitting $\Delta_{CEF}$, and the on-site Coulomb repulsion $U$. This could lead to a new iron oxidation state or a shift from the high-spin to low-spin configuration. However, such a shift would be visible as a shift in spectral weight of the X-ray absorption spectrum, which our results do not support. Still, a spin reorientation could occur through the presence of an unknown additional magnetic exchange coupling.

\subsection{A Frustrated, Glassy Phase}
\par Instead of another long-range ordered state, it is possible that our LSFO thin films transition into a low temperature glassy state similar to that seen in LSFO nanoparticles \cite{sabyasachiGlassyMagneticPhase2012}. In addition to supporting dimensionally-driven order destabilization, the nanoparticle study also creates precedent that finite size can drive the ordered state into a frustrated low temperature regime. Only in the smallest particles ($\sim70$ nm) did the study observe magnetic phase separation together with a cluster-glass-like transition near 65 K \cite{sabyasachiGlassyMagneticPhase2012}. This suggests that reducing the characteristic length scale in LSFO does not only suppress the bulk CD/AFM phase but can also promote magnetic frustration at the domain walls once long-range coherence is lost. Therefore, the increase in FC-ZFC splitting in our films could be explained by frustrated freezing at low temperature after the breakdown of the long-range ordered state. Recent work may identify the source of this frustration as the existence of multiple allowable charge order configurations \cite{nguyenMetastableChargeOrders2025}. Note that these calculations are about charge order rather than magnetic order alone and thus do not explicitly prove a rugged magnetic free-energy landscape; however, due to the strongly coupled orders of LSFO, one may infer that multiple coupled configurations in the free energy landscape may exist and produce frustration in the magnetic order.

\par Frustrated or glassy phases are typically characterized by slower dynamics, but in contrast, our coherent X-ray scattering measurements show that LSFO's magnetic domain dynamics instead speed up at low temperatures (Fig. \ref{fig:FIG5}c, d). It is important to note that these scattering measurements require the $\vec{q}=$\onesixth\ AFM Bragg peak to still be measurable, so they can only probe how the early onset of the low temperature phase impacts the $\vec{q}=$\onesixth\ AFM order. In the case of a glassy system, faster dynamics could be interpreted as enhanced motion of short-range magnetic domains or domain walls as long-range coherence is lost before any final frozen state emerges. A useful dynamical analogy is provided by La$_{1.72}$Sr$_{0.28}$NiO$_4$ (LSNO), where XPCS studies of both spin and charge stripes found that the low temperature regime is accompanied by faster dynamics and reduced spatial correlations \cite{ricciLSNOSpinDynamics2021, campiLSNOChargeDynamics2022}. The spin density wave study in LSNO reported an anomalous low-T speedup of spin stripe fluctuations as the stripe correlations weakened \cite{ricciLSNOSpinDynamics2021}. Further, the LSNO charge density wave study found that short correlation length charge puddles are more dynamic than the larger quasi-static region \cite{campiLSNOChargeDynamics2022}. Although LSNO does not imply the same microscopic mechanism as LSFO, it supports the broader interpretation that the loss of correlation length can coincide with faster local dynamics, while the eventual lower-temperature outcome may still be a frustrated or glassy state.

\section{Conclusion}
\par Cumulatively, these results support a picture in which fragility of long-range magnetic order drives a finite-size destabilization of the coupled CD/AFM phase in LSFO thin films. The system first enters the conventional long-range ordered state below $T_N$, but upon further cooling below $T^*\sim$60 K, that state can no longer be sustained, leading to collapse of both order parameters without an accompanying valence change or the appearance of a new diffraction peak. This work also unveils fast dynamics at the onset of order destabilization. Determining the identity of the low temperature phase remains a significant and interesting open question, as this work may suggest the existence of a lower energy AFM ground state or a complex, phase-separated glassy state. Differentiation of these possibilities will require probes that remain sensitive after the Bragg peaks have disappeared, such as magnetic neutron scattering or muon spin relaxation. Given the critical role of coupled orders and phase separation in phenomena such as colossal magnetoresistance and high-temperature superconductivity \cite{dagottoColossalMagnetoresistance2001,dagottoPhaseSeparation2003}, this work and its continuation is essential to better understanding ground states and order-disorder transitions in perovskite oxide systems. Furthermore, if the low-temperature phase is verified to be a spin or cluster glass, the presence of an AFM-to-glassy transition in a perovskite thin film could open the door for new spintronic and neuromorphic devices, offering the potential to switch between the spin-pumping capabilities in AFM/FM bilayers and the memory effects present in glassy systems \cite{grollier2020,liao2023}.
\section{Methods}
\par LSFO thin films were grown by pulsed laser deposition using a KrF laser ($\lambda$ = 248 nm) with a deposition rate of 4 Hz. The deposition was conducted at 710\textdegree C in an oxygen partial pressure of 100 mTorr. Following the deposition, the films were annealed in oxygen in-situ at a pressure of 10 Torr for 10 min. Films were grown on perovskite substrates SrTiO$_3$ (STO) in (111) and (100) orientations and on (LaAlO$_3$)$_{0.3}$(Sr$_2$TaAlO$_6$)$_{0.7}$ (LSAT) with (100) and (111) orientations. X-ray reflectivity and X-ray diffraction measurements were conducted with a Rigaku Smartlab X-ray diffractometer and a Huber 4-Circle X-ray diffractometer to verify film thickness and quality.
\par Atomic-resolution STEM-HAADF imaging was performed using a JEOL ARM200F transmission electron microscope operated at 200 kV with a convergence semi-angle of 21.2 mrad. Cross-sectional TEM specimens were prepared by focused ion beam (FIB) milling. Specimen thickness in the analyzed regions was estimated to be approximately 0.5$\lambda$, where $\lambda$ is the inelastic mean free path determined by the EELS log-ratio method.

\par Resonant elastic soft X-ray scattering (RSXS) experiments were performed at the Advanced Photon Source (APS) beamline 29-ID-C and the Stanford Synchrotron Radiation Lightsource (SSRL) beamline 13-3. All RSXS measurements were performed at the Fe $L_3$ edge (710 eV) with $\pi$-incident polarization and a horizontal scattering geometry (Fig. \ref{fig:result}a). The exact incident energy for all presented resonant measurements was chosen to be the resonance maximum for each unique measurement/beamline/sample geometry combination. Temperature-dependent scattering was taken around the Fe $L_3$ edge by scanning along the [111]$_{\textrm{pc}}$ direction while varying the temperature. In addition, resonant elastic hard X-ray scattering (REXS) experiments were conducted at the National Synchrotron Light Source II (NSLS II) at beamline 4-ID at the Fe $K$ edge (7.128 keV) to study the charge order at $\vec{q}=$\onethird. Measurements were taken with $\sigma-$incident polarization in a vertical scattering geometry. Energy dependence of the magnetic and charge ordered peaks was taken at a fixed Q.
\par  X-ray Magnetic Circular Dichroism (XMCD) was performed on a 35 nm sample of LSFO grown on STO(100) at the Advanced Light Source (ALS) beamline 4.0.2 vector magnet endstation. Electron yield (EY) was acquired while scanning around the Fe $L_{2, 3}$ edges under a fixed magnetic field of $\pm$0.1 T with both right- and left-helically polarized X-rays. Four measurements for each combination of field and polarization were collected. The X-rays were directed onto the sample with 20\textdegree\ grazing incidence, and the magnetic field was applied parallel to the incident beam direction. At each temperature step, the sample was left to equilibrate for 30 minutes before taking measurements. Reported XMCD spectra are produced by first edge normalizing the raw EY spectra, then via the subtraction:  
$$\mathrm{Avg}[(RHP - LHP)]_{+H} - \mathrm{Avg}[(RHP - LHP)]_{-H}$$
\par  X-ray absorption spectroscopy (XAS) was performed on a 30 nm sample of LSFO grown on LSAT(111) at the Advanced Light Source (ALS) beamline 4.0.2 vector magnet endstation. Luminescence yield (LY) was acquired while scanning around the Fe $L_{2, 3}$ edges under a fixed magnetic field of 0.45 T with both right- and left-helically polarized X-rays. The X-rays were directed onto the sample with 20\textdegree\ grazing incidence, and the magnetic field was applied parallel to the incident beam direction. At each temperature step, the sample was left to equilibrate for 30 minutes before taking measurements. Reported XAS spectra are an average of the LY spectra from the left- and right-circularly polarized incident beam.

\par Magnetometry was taken using a Quantum Design Physical Property Measurement System (PPMS) with a vibrating sample magnetometer (VSM). For field-cooled (FC) and zero field-cooled (ZFC) measurements, the sample was cooled in 100 Oe and 0 Oe respectively, and subsequently measured in a field of 100 Oe. Data were averaged over 5 measurements to reduce noise. Resistivity measurements were also taken in the PPMS under 0 Oe, with contacts bonded to the sample using silver paint.

\par Coherent scattering of the $\vec{q}=$\onesixth AFM order peak was performed at the Advanced Light Source (ALS) beamline 7.0.1.1 using the COSMIC endstation with a Princeton MTE3 CCD detector at a sample-to-detector distance of 0.208 m and a pixel size of 15$\times$15$\mu$m. Measurements were taken with a vertical scattering geometry and $\pi$-incident polarization at the Fe $L_3$ edge (712.5 eV). The beam was focused and filtered with a pinhole to a coherent spot size of 7 $\mu$m. This size was optimized such that 1 speckle covered $\sim$4 pixels. At each temperature step, the sample was left for 25 min to equilibrate before images of the AFM peak were collected.

\section{Acknowledgments}
\par The authors acknowledge fruitful discussions with Yayoi Takamura and Shriram Ramanathan. This work was supported as part of Quantum Materials for Energy Efficient Neuromorphic Computing (Q-MEEN-C), an Energy Frontier Research Center funded by the U.S. Department of Energy (DOE), Office of Science, Basic Energy Sciences (BES), under Award \# DE-SC0019273. This research used resources of the Advanced Photon Source, a U.S. Department of Energy (DOE) Office of Science user facility operated for the DOE Office of Science by Argonne National Laboratory under Contract No. DE-AC02-06CH11357. This research used beamline 4-ID of the National Synchrotron Light Source II, a U.S. Department of Energy (DOE) Office of Science User Facility operated for the DOE Office of Science by Brookhaven National Laboratory under Contract No. DE-SC0012704. This research used resources of the Advanced Light Source, a U.S. DOE Office of Science User Facility under contract no. DE-AC02-05CH11231. Use of the Stanford Synchrotron Radiation Lightsource, SLAC National Accelerator Laboratory, is supported by the U.S. Department of Energy, Office of Science, Office of Basic Energy Sciences under Contract No. DE-AC02-76SF00515.
\bibliography{references}

\clearpage

\setcounter{figure}{0}
\renewcommand{\thefigure}{S\arabic{figure}}

\setcounter{table}{0}
\renewcommand{\thetable}{S\arabic{table}}

\setcounter{equation}{0}
\renewcommand{\theequation}{S\arabic{equation}}
\setcounter{section}{0}
\renewcommand{\thesection}{S\arabic{section}}

\begin{center}
{\Large\bfseries Supplementary Information}
\end{center}

\section{Sample Properties}

\begin{figure*}[ht]
  \centering
  \includegraphics[width=\textwidth]{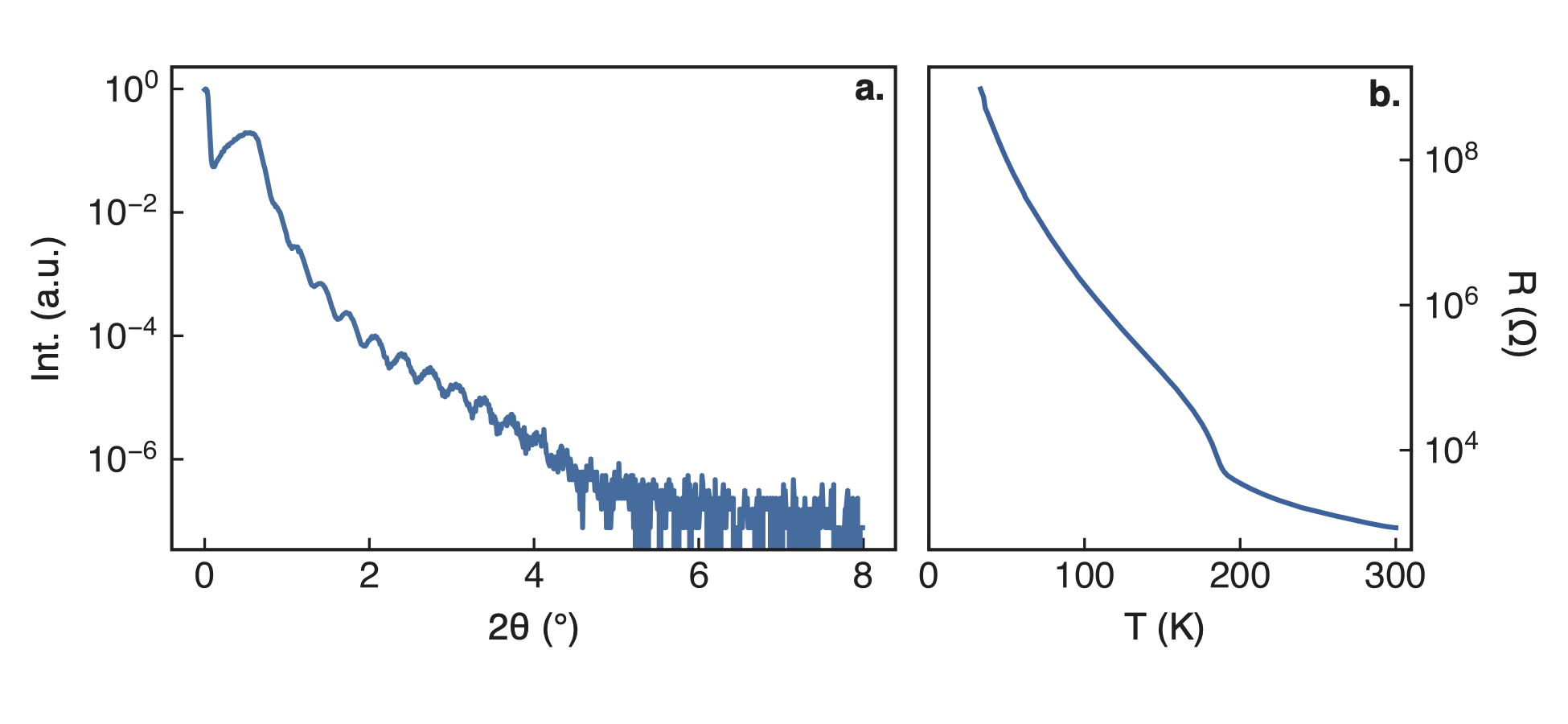}
  \caption{\textbf{Thickness and Resistance Characterization.} a) Example X-ray reflectivity for LSFO thin films. b) Resistance of LSFO thin film as a function of temperature.}
  \label{fig:SI1}
\end{figure*}
\lsfo\ film thicknesses were measured by X-ray reflectivity using Cu K$\alpha_1$ radiation ($\lambda = 1.54056 \mathring{\textrm{ A}}$). The reflectivity curve shows Kiessig fringes that come from the interference of the X-rays reflected from film and substrate surfaces (Fig. \ref{fig:SI1}a). The angular positions of the fringe peaks were identified and plugged into the equation below:
\begin{equation}
    \left(\frac{n\lambda}{2d}\right)^2=\textrm{sin}^2\theta_n-\textrm{sin}^2\theta_c
\end{equation}
where $n$ is the order of fringes, $\theta_n$ is the corresponding angle, and $\theta_c$ is the critical angle. By performing a linear fit of $n^2$ and $\textrm{sin}^2\theta_n$ one can extract the film thickness $d$ from the fitted slope.
\par Due to equipment failure, XRR was not obtained for some samples. When this was the  case, the thickness of the film was estimated using the FWHM of the film Bragg reflection from X-ray diffraction data. Presented below is a table containing the estimated thicknesses for all films studied in this report.
\def\arraystretch{1.25}
\begin{center}
\begin{tabular}{|c|c|c|c|c|} 
 \hline
 Substrate & Thickness &  Figure\\
 \hline
 LSAT(100) & $26.6\pm0.2$ nm & 1c, S1a\\
 \hline
 STO(111) & $46\pm4$ nm & 2c-e, 3d, 5\\ 
 \hline
 STO(111) &  $29\pm5$ nm & 2g-i\\ 
 \hline
 
 STO(100) & $14\pm2$ nm & 3e\\
 \hline
 LSAT(100) & $39.3\pm0.2$ nm & 3f, 4d-e, S1b\\
 \hline
 STO(100) & $35.5 \pm 1.5$ nm  & 4a\\ 

 \hline
 LSAT(111) & $30\pm1$ nm & 4b-c\\
 \hline
\end{tabular}
\end{center}

\section{Additional Scattering Data}
\begin{figure*}[h]
  \centering
  \includegraphics[width=\textwidth]{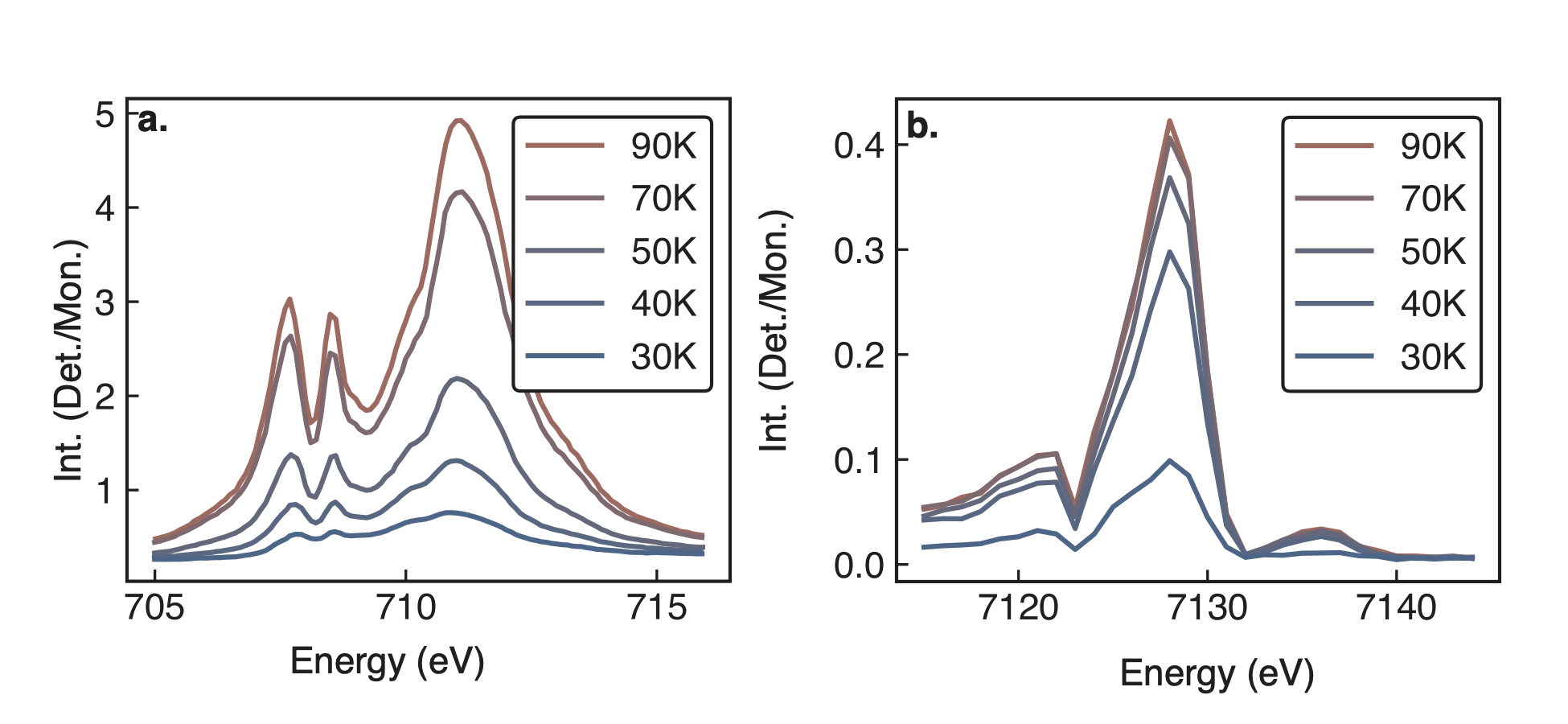}
  \caption{\textbf{Resonant Scattering Energy Dependence.} a) Energy dependence of the magnetic order peak at $\vec{q}=$\onesixth\ for LSFO grown on STO(100), taken at a fixed Q. b) Energy dependence of the charge order peak at $\vec{q}=$\onethird\ for LSFO grown on STO(111), taken at a fixed Q. }
  \label{fig:SI2}
\end{figure*}

\par In addition to temperature-dependent scattering, energy dependencies of the $\vec{q}=$\onesixth\ AFM and $\vec{q}=$\onethird\ CD peaks were taken (Fig. \ref{fig:SI2}a,b respectively). These  data were collected by aligning to the ordering peak and measuring the scattered intensity at fixed Q while varying the incident X-ray energy. Sharp resonances are seen around the Fe $L_3$ edge for the AFM peak and around the Fe $K$ edge for the CD peak.

\begin{figure*}[h]
  \centering
  \includegraphics[width=\textwidth]{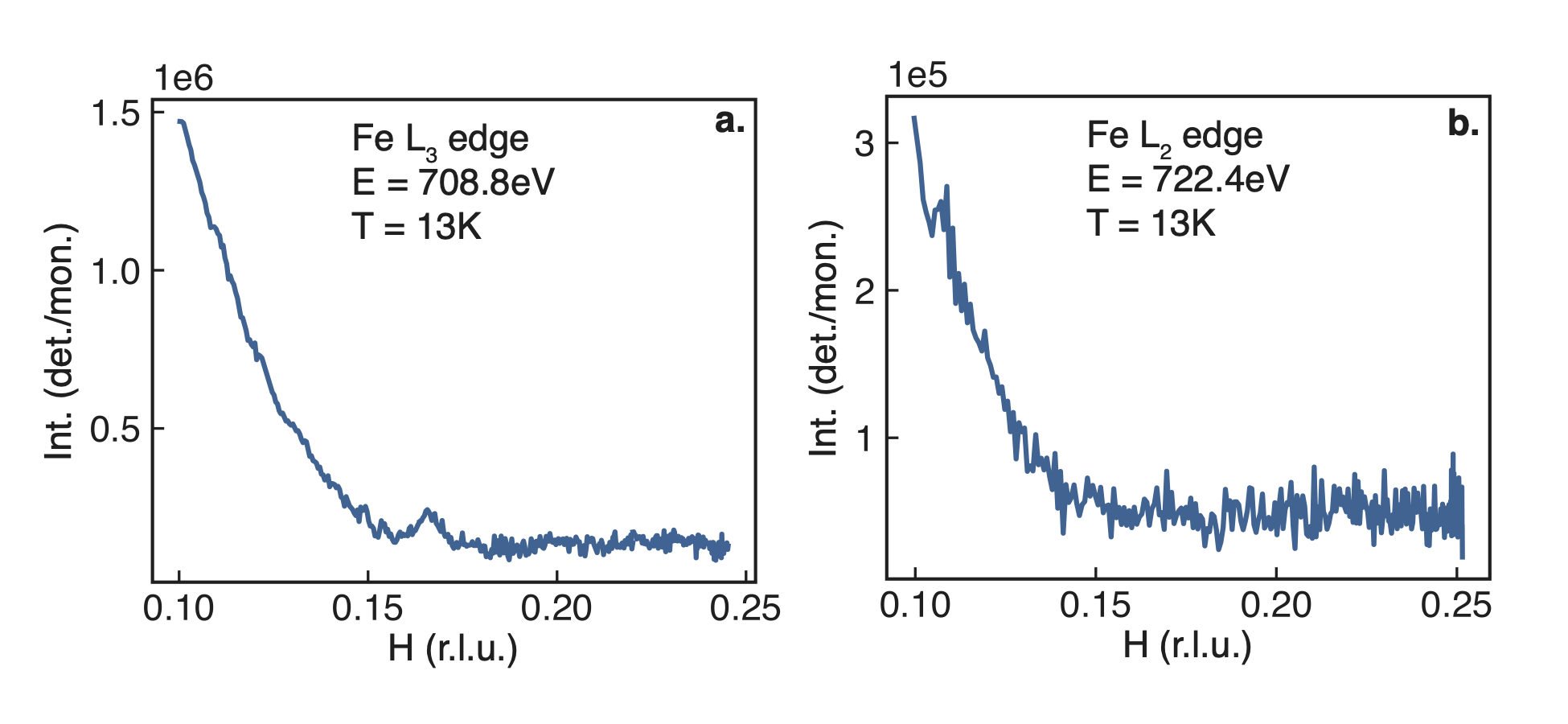}
  \caption{\textbf{RSXS Search for New Peaks.} a) Scan along the [111]$_{\textrm{pc}}$  direction taken at $T=13$ K at the Fe $L_3$ edge. No additional magnetic ordering peaks are visible. b) The same scan taken at the Fe $L_2$ edge for increased reciprocal space access. Again, no additional peaks are seen.}
  \label{fig:SI3}
\end{figure*}

\par Peak searches were conducted along the [111]$_{\textrm{pc}}$ direction with RSXS to search for evidence of a new AFM order at low temperatures. Due to limited angular access, scans were only possible from $H=0.1-0.245$ at the Fe $L_3$ edge (Fig. \ref{fig:SI3}a). In this scan, only the $\vec{q}=$\onesixth\ magnetic peak is visible. Scans were also taken along the higher-energy Fe $L_2$ edge (Fig. \ref{fig:SI3}b) to access a slightly larger range in $H$, though the $\vec{q}=$\onesixth\ ordering peak is less prominent at this edge. Again, no additional evidence of new order is seen in this range.

\section{Coherent X-ray Scattering}
\begin{figure*}[ht]
  \centering
  \includegraphics[width=\textwidth]{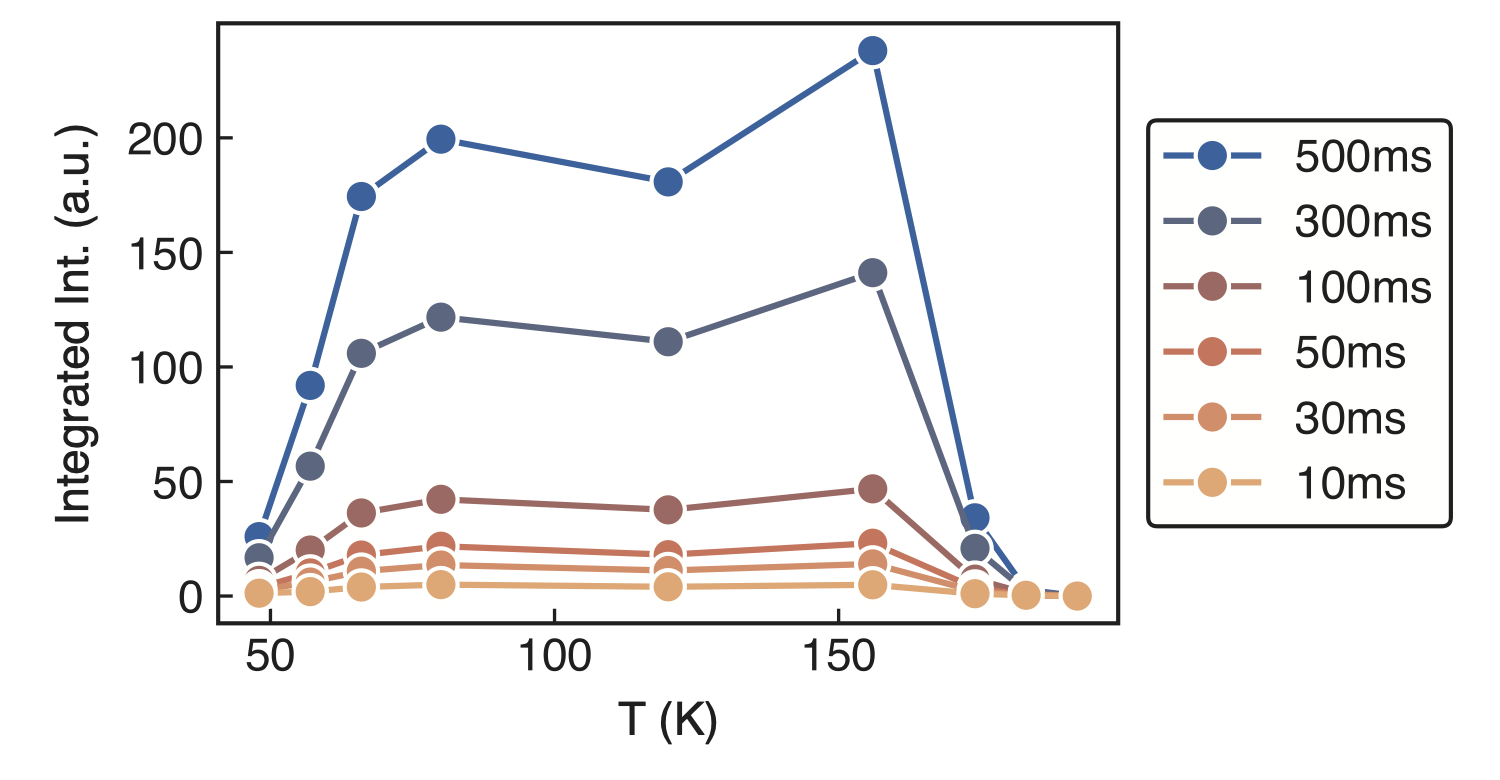}
  \caption{Temperature dependence of magnetic order peaks extracted from XSVS speckle patterns.}
  \label{fig:SI4}
\end{figure*}
During data collection for the time series corresponding to Fig. 5d and 5b, images were taken with a readout time of 725.34 ms. Fig. \ref{fig:SI4} shows the integrated peak intensity for the same images $\beta$ was calculated for in Fig. 5d. A significant change in contrast was observable even for temperatures with nearly equivalent integrated peak intensities, indicating that the change in contrast cannot be attributed to a simple change in signal-to-noise ratio as the peak intensity drops. 

\par For the time series corresponding to Fig. 5c, the readout time used was 318.4 ms, and the exposure time was optimized for best signal intensity at a given temperature. At 175 K, an exposure time of 200 ms was used, and for all other temperatures a 500 ms exposure time was used. 

\begin{figure*}[ht]
  \centering
  \includegraphics[width=\textwidth]{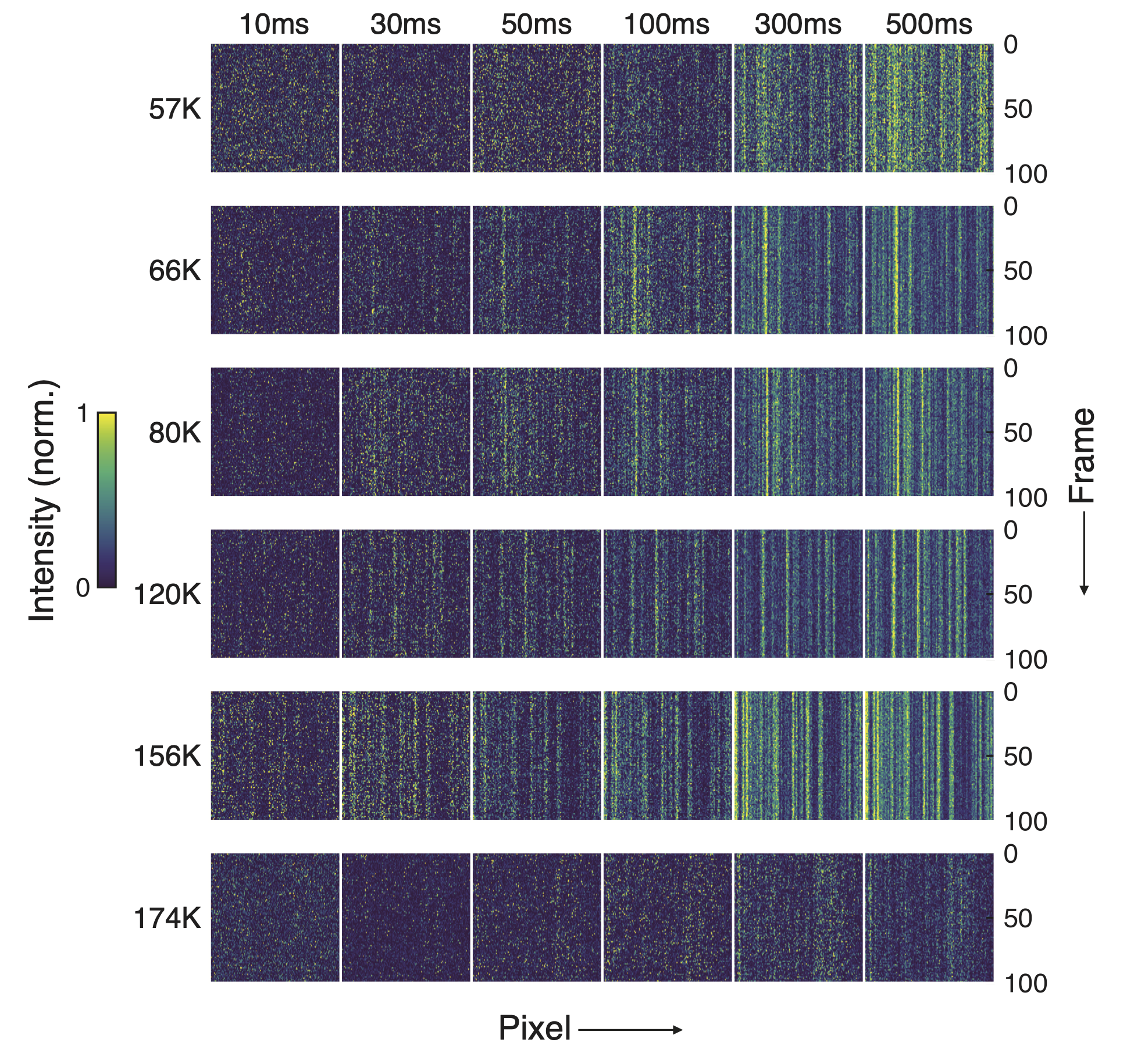}
  \caption{Waterfall plots for speckle patterns taken as a function of exposure time and temperature.}
  \label{fig:SI5}
\end{figure*}

The raw data was dark and flat field corrected pixel by pixel using
$$C = \frac{(R-D)\times m}{(F-D)} = (R-D)\times G$$

\noindent where:
    $C$ = corrected image\\
    $R$ = raw image\\
    $F$ = flat field image\\
    $D$ = dark frame image\\
    $m$ = image-averaged value of $(F-D)$\\
    $G$ = gain = $\frac{m}{(F-D)}$\\

There is still some distribution of pixel intensities below zero but these values are from noise. Therefore, corrected images were thresholded below zero since everything above should correspond to real photon counts from speckles.

The second order correlation, $g_2$, given by 
$$g_2(q, \Delta t) = 1+\beta\left|g_1(q, \Delta t)\right|^2$$
was calculated over an ROI on the $\vec{q}=$\onesixth\ peak just off the center. This was necessary to avoid including contributions from the truncation rod, since in the (111)-oriented samples, the magnetic order is specular. From the $g_2$, we extract and compare the contrast from $g_2(\Delta t=\text{exposure time})$. Note that it would be incorrect to use $\Delta t=0$, as this is the self correlation. 

\end{document}